\documentclass[10pt,conference]{IEEEtran}
\IEEEoverridecommandlockouts

\usepackage{cite}
\usepackage{array,framed}
\usepackage{booktabs}
\usepackage{
  float,
  epsfig,
  wrapfig,
  graphics,
  graphicx,
  subcaption
}
\usepackage{xparse}
\newcommand{\bnm}{\begin{newmath}}
\newcommand{\enm}{\end{newmath}}

\newcommand{\bea}{\begin{eqnarray*}}%
\newcommand{\eea}{\end{eqnarray*}}%

\newcommand{\bne}{\begin{newequation}}
\newcommand{\ene}{\end{newequation}}

\newcommand{\bal}{\begin{newalign}}
\newcommand{\eal}{\end{newalign}}

\newenvironment{newalign}{\begin{align}%
\setlength{\abovedisplayskip}{4pt}%
\setlength{\belowdisplayskip}{4pt}%
\setlength{\abovedisplayshortskip}{6pt}%
\setlength{\belowdisplayshortskip}{6pt} }{\end{align}}

\newenvironment{newmath}{\begin{displaymath}%
\setlength{\abovedisplayskip}{4pt}%
\setlength{\belowdisplayskip}{4pt}%
\setlength{\abovedisplayshortskip}{6pt}%
\setlength{\belowdisplayshortskip}{6pt} }{\end{displaymath}}

\newenvironment{newequation}{\begin{equation}%
\setlength{\abovedisplayskip}{4pt}%
\setlength{\belowdisplayskip}{4pt}%
\setlength{\abovedisplayshortskip}{6pt}%
\setlength{\belowdisplayshortskip}{6pt} }{\end{equation}}

\newcounter{ctr}

\newcounter{mytable}
\def\mytable{\begin{centering}\refstepcounter{mytable}}
\def\endmytable{\end{centering}}

\newcounter{myfig}
\def\myfig{\begin{centering}\refstepcounter{myfig}}
\def\endmyfig{\end{centering}}

\newlength{\saveparindent}
\newlength{\saveparskip}
\newcommand{\E}{{\rm I\kern-.3em E}}

\def \part {part}

\renewcommand{\paragraph}[1]{\vspace*{6pt}\noindent\textbf{#1}\;}

\def \blackslug{\hbox{\hskip 1pt \vrule width 4pt height 8pt
    depth 1.5pt \hskip 1pt}}
\def \qed{\quad\blackslug\lower 8.5pt\null\par}

\newcounter{mynote}[section]

\newcommand\ignore[1]{}

\newcounter{rcnote}[section]

\newcounter{mrnote}[section]

\newcounter{fknote}[section]

\newcounter{anote}[section]

\DeclareMathSymbol{\mlq}{\mathord}{operators}{``}
\DeclareMathSymbol{\mrq}{\mathord}{operators}{`'}

\newcommand{\rhf}[2]{R_{f, \gamma}}

\DeclareDocumentCommand{\edist}{o o}{
  \ensuremath{
    \IfNoValueTF{#1}{{d}}{{\sf d}(#1,#2)}
  }
}

\newcommand{\olrk}[1]{\ifx\nursymbol#1\else\!\!\mskip4.5mu plus 0.5mu\left(\mskip0.5mu plus0.5mu #1\mskip1.5mu plus0.5mu \right)\fi}

\NewDocumentCommand{\indseq}{ O{1} O{r} }{{#1}\ldots {#2}}

\usepackage{listings}
\lstdefinestyle{Python}{
    language        =   Python, 
    basicstyle      =   \zihao{-5}\ttfamily,
    numberstyle     =   \zihao{-5}\ttfamily,
    keywordstyle    =   \color{blue},
    keywordstyle    =   [2] \color{teal},
    stringstyle     =   \color{magenta},
    commentstyle    =   \color{red}\ttfamily,
    breaklines      =   true,   
    columns         =   fixed,  
    basewidth       =   0.5em,
}
 
 \usepackage{enumitem}
\usepackage{textcomp,amssymb}
\usepackage{setspace}
\usepackage{amsfonts}
\usepackage{latexsym,fancyhdr}
\usepackage{enumerate}
\usepackage{url}
\usepackage{algpseudocode}
\usepackage{graphics}
\usepackage{xparse} 
\usepackage{xspace}
\usepackage{multirow}
\usepackage{csvsimple}
\usepackage{balance}
\usepackage{listings}
\usepackage{
  tikz,
  pgfplots,
  pgfplotstable
}
\usepackage{hyperref}
\usetikzlibrary{
  shapes.geometric,
  arrows,
  external,
  pgfplots.groupplots,
  matrix
}
\pgfplotsset{compat=1.9}
\usepackage{mathtools}

\DeclareMathAlphabet{\mathcal}{OMS}{cmsy}{m}{n}

\DeclareGraphicsExtensions{%
    .png,.PNG,%
    .pdf,.PDF,%
    .jpg,.mps,.jpeg,.jbig2,.jb2,.JPG,.JPEG,.JBIG2,.JB2}

\definecolor{darkblue}{rgb}{0, 0, 0.5}
\hypersetup{
	colorlinks=true,
	linkcolor=black,
	citecolor=darkblue,
	filecolor=black,
	urlcolor=darkblue,
	pdfauthor = {},
	pdftitle = {},
	pdfkeywords = {}
}
\usepackage{threeparttable}
\usepackage{amsmath}
\usepackage{colortbl}
\usepackage{booktabs}
\usepackage{graphicx}
\usepackage{xcolor}
\usepackage[linesnumbered,ruled,vlined]{algorithm2e}

\SetCommentSty{mycommfont}
\SetKwInput{KwInput}{Input}                
\SetKwInput{KwOutput}{Output} 

\newcommand{\smallbf}[1]{{\smaller{\textbf{#1}}}}

\usepackage{tcolorbox}
\newcommand{\answer}[2] {
    \begin{tcolorbox}[boxrule=0.5pt,left=1pt,right=1pt,top=1pt,bottom=1pt]
        \smallbf{Answer to RQ#1:} #2
    \end{tcolorbox}
}

\newcommand{\oursystem}{{ATPChecker}}
\newcommand{\soot}{\textit{soot}}

\newcommand{\maven}{{Maven Repository}}

\newcommand{\app}{{app}}
\newcommand{\appbrain}{{AppBrain}}
\newcommand{\tpl}{{TPL}}
\newcommand{\github}{\textit{GitHub}}

\definecolor{turquoise}{rgb}{0.1, 0.75, 0.6}
\usepackage{pifont}

\newcommand{\kaifa}[1]{{\textcolor{black}{#1}}}

\def\BibTeX{{\rm B\kern-.05em{\sc i\kern-.025em b}\kern-.08em
    T\kern-.1667em\lower.7ex\hbox{E}\kern-.125emX}}

\begin{document}


\title{Demystifying Privacy Policy \\ of Third-Party Libraries in Mobile Apps}

\author{ \rm 
\IEEEauthorblockN{
Kaifa Zhao\IEEEauthorrefmark{2}, 
Xian Zhan\IEEEauthorrefmark{3}, 
Le Yu\IEEEauthorrefmark{2}\thanks{\IEEEauthorrefmark{1} The corresponding authors.}\IEEEauthorrefmark{1}, 
Shiyao Zhou\IEEEauthorrefmark{2}, 
Hao Zhou\IEEEauthorrefmark{2}, 
Xiapu Luo\IEEEauthorrefmark{2},
Haoyu Wang\IEEEauthorrefmark{4}, 
Yepang Liu\IEEEauthorrefmark{3}}
\IEEEauthorblockA{
{\IEEEauthorrefmark{2} The Hong Kong Polytechnic University, 
\IEEEauthorrefmark{3} Southern University of Science and Technology}, 
\\
{\IEEEauthorrefmark{4} Huazhong University of Science and Technology}}
\IEEEauthorblockA{
\textit{
{kaifa.zhao@connect.polyu.hk}, 
chichoxian@gmail.com, 
yulele08@gmail.com, 
shiyao.zhou@connect.polyu.hk,
}}
\IEEEauthorblockA{\textit{
cshaoz@comp.polyu.edu.hk, 
{csxluo@comp.polyu.edu.hk}, 
haoyuwang@hust.edu.cn, 
liuyp1@sustech.edu.cn}}
}

\maketitle
\sloppy
\thispagestyle{plain}
\pagestyle{plain}

\begin{abstract}

The privacy of personal information has received significant attention in mobile software.
Although researchers have designed methods to identify the conflict between app behavior and privacy policies, little is known about the privacy compliance issues  relevant to  third-party libraries (TPLs).
The regulators enacted articles to regulate the usage of personal information for TPLs (e.g., the 
CCPA requires businesses clearly notify consumers if they share consumers' data with third parties or not).
However, it remains challenging to investigate the privacy compliance issues of TPLs due to three reasons:
1) Difficulties in collecting TPLs' privacy policies. 
%
%
%
In contrast to Android apps, which are distributed through markets like Google Play and must provide privacy policies, there is no unique platform for collecting privacy policies of TPLs.
2) Difficulties in analyzing TPL's user privacy access behaviors.
%
TPLs are mainly provided in binary files, such as \textit{jar} or \textit{aar}, and their whole functionalities usually cannot be executed independently without host apps.
3) Difficulties in identifying consistency between TPL's functionalities and privacy policies, and host app's privacy policy and data sharing with TPLs.
This requires analyzing not only the privacy policies of TPLs and host apps but also their functionalities.
%
In this paper, we propose an automated system named {\oursystem} to analyze whether Android TPLs comply with the privacy-related regulations.
%
We construct a data set that contains a list of 458 TPLs, 247 TPL's privacy policies, 187 TPL's binary files
and 641 host apps and their privacy policies. 
Then, we analyze the bytecode of TPLs and host apps, design natural language processing systems to analyze privacy policies, and implement an expert system to identify TPL usage-related regulation compliance.
The experimental results show that 31\% TPLs violate regulation requirements for providing privacy policies. 
Over 39.5\% TPLs miss disclosing data usage in their privacy policies. 
Over 47\% host apps share user data with TPLs and over 20\% of host apps miss disclosing usage of TPLs and interactions with TPLs.
Our findings remind developers to be mindful of TPL usage when developing apps or writing privacy policies to avoid violating regulations.

\end{abstract}


\section{Introduction}

Nowadays, smartphones and mobile apps are playing essential roles in our daily lives~\cite{deepintent2019SIGSAC}. 
More and more developers tend to use many off-the-shelf third-party libraries (TPLs) to facilitate the development process, to avoid reinventing the wheel~\cite{atvhunter2021ICSE}.
However, developers may omit the privacy threat in TPLs. 
For example, the advertisement libraries may leak users' personal information (e.g., IMEI)~\cite{short2014android}.
Some TPLs (e.g., advertising and analytic libraries) rely on personal information to provide better service~\cite{demetriou2016free}.
However, TPLs may collect user data that is not necessary for providing related services. 
Such behaviors may lead to privacy leakage or even security concerns~\cite{harty2021logging, tangtse:21a, zhao2021structural}.
Besides, the permission mechanism of Android also increases the possibility of leaking user privacy via TPLs~\cite{lin2012expectation, kawabata2013sanadbox, zhang2013aframe, liu2015efficient, shekhar2012adsplit, pearce2012addroid,he2018investigation, salza2018developers, zhan2017splitting, wang2014compac, sun2014nativeguard}.
Since the host apps share the same process space and permissions with TPLs, 
the mechanism violates the principle of least privilege, which leads to TPLs being over-privileged.
Recent studies find that permission abuse allows TPLs to access users' personal information and causes potential privacy leakage~\cite{qiu2021libcapsule, grace2012unsafe, birendra2016android,xu2022lie,li2021embedding,di2021influencing,seneviratne2015measurement,cybernews_dataleakage}. 
In addition, the incorrect use of TPLs~\cite{zhang2020does, chen2014information, cybernews_dataleakage} may cause serious personal information leakage. 
For example, a recent report~\cite{cybernews_dataleakage} discloses that popular Android apps with over 142.5 million installations leak user data to unauthorized third parties. 
Zhang et al.~\cite{zhang2020does} disclose 120 of 555 apps that share users' personal information to analytic service TPLs without encryption. 
Besides, linking and mining large amounts of unencrypted personal information probably pose threats to serious personal privacy leakage~\cite{kiranmayi2018reducing, chen2014information}.
Thus, it is urgent to design a system to check the data access behavior and privacy policies of TPLs so that we can determine if TPLs are compliant with privacy regulations.

Various privacy-related regulations 
(e.g., {\textsc{\small{California Consumer Privacy Act (CCPA), General Data Protection Regulation (GDPR)}}})~\cite{piss, zhSDKLaw, gdpr, ccpa, cfr})
have been promulgated to protect people's personal information from being abused.
However, app developers may unconsciously violate regulations by using certain TPLs because they may just follow the user guidelines to invoke these TPLs without knowing the internals of these TPLs. 
Moreover, due to a lack of security consciousness, developers may never check whether the TPLs used in their apps have a privacy policy and whether these TPLs may have security risks with respect to privacy leakage. 
%
The latest \textit{\small{Google Play Developer Program policies (GPDP)}}~\cite{google_play_pp_update} asks developers to ensure the TPLs used in their apps are compliant with GPDP~\cite{gpdp}.
Developers of {\tpl}s are required to provide privacy policies and disclose data usage.
Otherwise, apps may violate regulation or market requirements and be removed from the app market~\cite{googleplay_requirements, appgallery_requirements}.
To help TPLs comply with regulations and help developers correctly use TPLs,  it is necessary to analyze the consistency between {\tpl}s' data usage and privacy policies.

Previous work~\cite{ppcheck,ppchecker,andow2019policylint, yu2016can,nguyen2021share,zhang2020does,wang2021understanding} has proposed methods to analyze privacy issues of apps, such as analyzing the consistency between apps' privacy policy descriptions and behaviors.
However, existing methods cannot identify the app's privacy issues related to the usage of TPLs inside apps because existing methods fail to identify TPLs functionalities or analyze TPL privacy policies.
PoliCheck~\cite{ppcheck} performs dynamic analysis to get traffic flow for analyzing the apps' data usage.
Then, it applies data and entity dependency tree analysis to extract data usage statements from the privacy policy, and designs rules to identify the conflicts.
PPChecker~\cite{ppchecker} checks the trustworthiness of apps' privacy policies and consistency between an app's behavior and its privacy policy without considering the latest regulations. 
%

Thus, there is a research gap with respect to discrepancy analysis between the privacy-related behaviors of TPLs and regulations.
To this end, we propose a tool, named \oursystem ~(\underline{\textbf{A}}utomated \underline{\textbf{T}}hird-party library \underline{\textbf{P}}rivacy compliance \underline{\textbf{Checker}}) to automatically identify whether the usages of in-app TPLs complies with privacy-related regulations (e.g., {GDPR Article I.(47), CFR 332.6}).
As regulation analysis requires expert knowledge of both legal and software engineering, we invited 
one expert and one senior researcher to read through related regulations~\cite{gdpr, ccpa, cpra, piss,petitcolas1883cryptographie,iccpr,zhSDKLaw} and summarize requirements for TPLs.  
They have conducted research on software engineering and privacy policy analysis for over seven years and two years, respectively.
{\oursystem} identifies the inconsistency between TPLs and regulation by analyzing TPLs' bytecode (\S\ref{sec:tpl_data}) and privacy policies (\S\ref{sec:tpl_pp}).
For example, {\oursystem} discovers whether TPLs correctly provide privacy policies.
{\oursystem} discloses whether host app's TPL usage comply with regulations(\S\ref{sec:host_app_data}, \S\ref{sec:host_app_pp}).

It is non-trivial to develop {\oursystem}. 
First, it is not straightforward to get TPLs and corresponding privacy policies.
In contrast to Android apps, which are distributed through markets like Google Play with their privacy policies, there is no central platform for collecting privacy policies of TPLs.
To counter this issue, we first construct a data set that contains a TPL repository and a host app repository.
The TPL repository contains a) a TPL list, which contains 458 TPLs crawled from {\appbrain}~\cite{appbrain}, b) 187 TPLs' binary files, which are in the format of \textit{jar} or \textit{aar}, from {\maven}~\cite{mvnrepository}, and c) 247 TPLs' privacy policies which are manually collected. 
The details of dataset components and relations are given in \S\ref{sec:eva}. 
The host app repository contains a) 641 distinct host apps which are collected from {Google Play} based on the list from {\appbrain} and b) the host app's privacy policies.
Second, it is not easy to analyze TPLs' user privacy access behavior because TPLs are mainly provided in the format of binary files, such as \textit{jar} and \textit{aar}, and TPLs' whole functions may not be executed independently without being triggered by host apps.
{\oursystem} performs static analysis~\cite{soot,arzt2014flowdroid} on TPLs and uses data flow analysis to trace TPLs' personal information usage without the need of TPLs' source code (\S\ref{sec:tpl_data}).
We exclude dynamic analysis since it cannot execute all possible paths in apps or TPLs~\cite{schindler2022privacy}.
Third, analyzing privacy policy is difficult because developers use various natural languages to describe the usage of personal information and TPLs.
To analyze data and TPL usage-related statements in privacy policies, {\oursystem} implements an expert system to extract abstract data usage patterns (\S\ref{sec:host_app_pp}) and investigates TPL usage compliance.

Overall, our contributions are summarized as follows:

\begin{itemize}[leftmargin=*]

\item We propose a novel system named {\oursystem} to analyze the compliance of Android TPLs. 
{\oursystem} uses static analysis to identify TPL's user data access behaviors and host apps' data interaction with TPLs, and uses natural language processing techniques to analyze TPLs' and host apps' privacy policies.
Combining the results of bytecode analysis and privacy policy analysis, {\oursystem} determines whether TPLs and usage of TPLs in host apps comply with the regulation.

\item To evaluate the performance of {\oursystem} and facilitate further research in this area, we construct a privacy dataset that includes 187 TPLs' binary files and their privacy policies, and 641 host apps and their privacy policies.
%
%
%

%
\item {\oursystem} discovers over that 31\% of TPLs miss providing privacy policies and 39\% of TPLs' privacy policies conceal data usage.
{\oursystem} finds that over 20\% of host apps violate the regulation requirements for clearly disclosing data interactions with TPLs.

\end{itemize}
 
\section{Background}

\subsection{Android Third-party Libraries}
\label{sec:bb_tps}

Android TPLs provide abundant functions (e.g., user data analysis and advertising recommendations), which can be re-used by developers in their apps to facilitate development progress. However, TPLs may introduce data leakage issues.
%
%
Due to Android's permission mechanism, TPLs share the same privileges with host apps~\cite{qiu2021libcapsule, grace2012unsafe, birendra2016android}.
TPLs may abuse permissions to access users' personally identifiable information without users' consent~\cite{xu2022lie, li2021embedding}, which results in personal information leakage~\cite{tang:ase19, lin2012expectation, kawabata2013sanadbox, zhang2013aframe, liu2015efficient, shekhar2012adsplit, pearce2012addroid,he2018investigation, salza2018developers, zhan2017splitting, wang2014compac, sun2014nativeguard}.

Android third-party libraries are generally available as binary files, such as \textit{*.jar} or \textit{*.aar}.
Developers mainly import binary files into their projects to use the TPLs without inspecting the TPLs' data usage, which results in the violation of regulation requirements. 
Furthermore, it is very time-consuming for developers to understand TPLs' data usage behavior. 
Thus, developers may not be able to clearly describe the TPLs' data usage in their privacy policies or may just provide links to TPLs' privacy policies~\cite{ppchecker}, thus making their apps and privacy policies violate regulations. 

\subsection{Privacy Policy}
\label{sec:bb_pp}

%
Privacy policies describe how the data controllers use, disclose, store, manage and share users' personal information~\cite{piss, gdpr, ccpa,zhSDKLaw}. 
Regulations require privacy policies to describe TPLs' data access behaviors clearly. 
Besides, the coverage of personal information~\cite{gdpr} is not limited to identical information, such as ID, but also includes any information that can be used to identify or infer a specific person~\cite{gdpr}.
To comply with regulation requirements, privacy policies should clearly claim the following information~\cite{le2015ppg}:
1) personal information,
2) the software's contact information,
3) the purpose of collecting personal information,
4) types of data shared with third parties
and 5) the rights that users have.

\subsection{Regulation Requirements for TPLs and Usage of TPLs}
\label{sec:bb_regulation_req}

Regulations have enacted Articles to normalize personal information usage by software. 
\textsc{\small{California Consumer Privacy Act (CCPA)}}~\cite{ccpa} \textsc{\small{{Article 1798.120}}} claims that ``\textit{a business should notify consumers if they sell consumers' personal information to third parties}". 
\textsc{\small{Cybersecurity Practices Guidelines–Security Guidelines for Using Software Development Kit (SDK) for Mobile Internet Applications (App) (SGSDK)}}~\cite{zhSDKLaw}, has been enacted to standardize the management of third-party libraries.
For example, \textsc{\small{SGSDK Article 5.1 d)}} claims ``\textit{The SDK discloses the scope, purpose, and rules of the SDK's processing of personal information to the App in a clear, understandable and reasonable manner. The actual behavior of the SDK's collection and use of personal information should be consistent with the statement in the public document.}''
Similar requirements are also mentioned in \textsc{\small{GDPR}}~\cite{gdpr} \textsc{\small{Article 14.2}} and \textsc{\small{CCPA}}~\cite{ccpa} \textsc{\small{Article 4}}. 
\textsc{\small{Information security technology-Personal information (PI) security specification (PISS)}}~\cite{piss} Article 9.7 specifies the requirements for third parties.
Regulations~\cite{MDICUPIA} also enact Articles to regulate usage of TPLs. 
For example, regulations~\cite{MDICUPIA,zhSDKLaw,piss,gdpr,cfr} require apps to disclose the purpose, method and scope of usage of TPLs.
%
%
In \textsc{PISS}, \textsc{GDPR} and \textsc{CCPA}, {\tpl}s are regarded as data controllers and are required to expose their data usage.

\section{Methodology}

\begin{figure*}[t]
    \centering
    \includegraphics[width=0.66\textwidth]{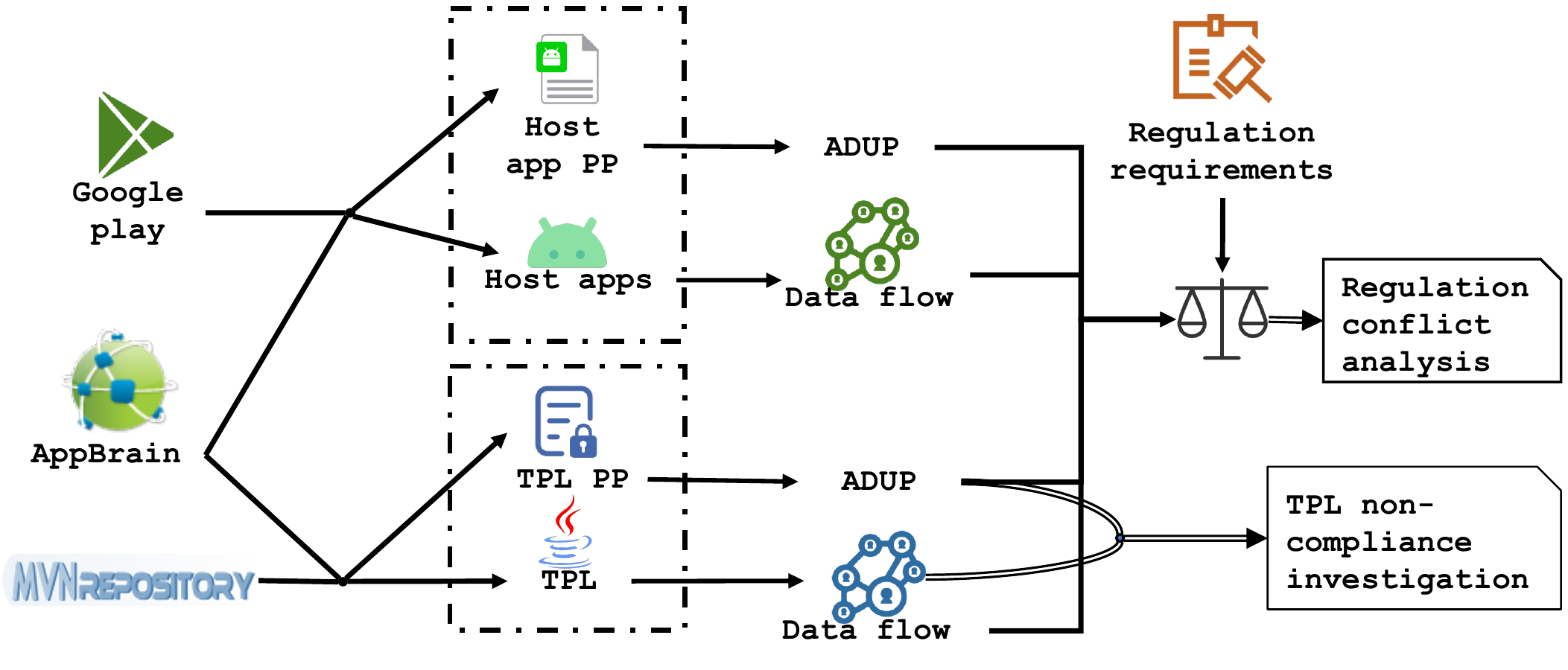}
    \vspace{0.5ex}
    \caption{Framework of {\oursystem}.} 
    \label{fig:sys_frame}
\end{figure*}

This section introduces how {\oursystem} identifies TPLs' data usage (\S\ref{sec:tpl_data}) and how to identify the consistency of TPLs' data usage with the statements in privacy policies (\S\ref{sec:tpl_pp}).
We will also describe how to determine whether host apps describe TPLs usage correctly in their privacy policies (\S\ref{sec:host_app_pp}) and whether host apps use TPLs correctly (\S\ref{sec:host_app_data}).
Fig.~\ref{fig:sys_frame} shows the framework of {\oursystem}.

\begin{table}[hb]
\centering
\smaller 
\vspace{1em}
\caption{Tracked data types in static analysis.}
\begin{tabular}{p{0.9\columnwidth}}
\toprule
\textbf{Data Type} \\ \midrule
Ad ID, username, password, name, location, contact, phone number, email address, IMEI, Wi-Fi, MAC address,  GSF ID, Android ID, serial number, SIM serial number \\ \bottomrule
\end{tabular}
\label{tab:data_type}
\end{table}

\subsection{Identify personal information usage in third-party libraries}
\label{sec:tpl_data}

{\oursystem} analyzes TPL's  personal information  (PI) usages by performing static analysis based on {\soot}~\cite{soot} and does not require TPLs' source code.
{\oursystem} conducts data flow analysis, specifically variable use-define analysis~\cite{defuse94}, to locate data of interest (DOI) (Tab.~\ref{tab:data_type}), which are mainly related to user identification information summarized from regulations. 
{\oursystem} uses function call graphs (FCG) to trace PI flow among methods. 
However, our analysis discovers that the soot cannot effectively discover the TPL's main functions and entry points to construct a valid FCG~\cite{arzt2014flowdroid}.
Thus, we propose the following methods to optimize the FCG construction and improve the data flow analysis precision.

\noindent\textbf{TPL FCG construction: }
{\oursystem} iterates over each class and method.
Since apks mainly use TPLs by invoking their \textit{public} methods, {\oursystem} extends TPLs \textit{public} methods to entry points set to optimize FCG construction.
%

\begin{algorithm}[t]
\caption{Data usage analysis}\label{alg:data_flow}
\label{alg:data_usage}
\SetKwInput{KwInput}{Input}                
\SetKwInput{KwOutput}{Output}              
\DontPrintSemicolon
  
  \KwInput{method $m_t$, statement $s_t$, variable $v_t$}
  \KwOutput{data usage $FD$ of variable $v_t$}

    \SetKwFunction{FInterMVA}{InterMethodVarAnalysis}
    \SetKwFunction{FIntraMVA}{IntraMethodVarAnalysis}
    \SetKwFunction{FForwar}{ForwardAnalysis}
    \SetKwFunction{FBackwar}{BackwardAnalysis}
  
    BackwardAnalysis($m_t$, $s_t$, $v_t$)\;

    IntraMethodVarAnalysis($m_t$, $s_t$, $v_t$) \;
    \SetKwProg{Fn}{Def}{:}{}
    \Fn{\FIntraMVA{$m$, $stmt$, $var$}}{
        $uses$ = find statements that use $var$ in method $m$ \;
        \For{$u$ in $uses$}{
            store($var$, u) in $DF$ \;
            ForwardAnalysis($m$, $u$, $var$)
        }
    }
    \;
    \SetKwProg{Fn}{Def}{:}{\KwRet}
    \Fn{\FBackwar{$m_t$, $s_t$, $v_t$}}{
        $defs$ = getDefsOfAt($v_t$, $s_t$, $cfg$)  \;
        \For{$d$ in $defs$}{
            $src_{var}$ = get the signature of $v_t$ in $d$\;
            $caller_{list}$ = find methods that invoke $m_t$ \;
            \For{$m$ in $caller_{list}$}{
                $src_{stmt}$ = statements that invoke $m_t$ in $m$\;
                $src_{v}$ = variables in $src_{stmt}$ that correspond to $v_t$\;
                \If{$src_{v}$  is \textit{Variable}}{
                    BackwardAnalysis($src_{v}$, $src_{stmt}$, $src_{m}$)
                } \ElseIf{$src_{v}$ is \textit{Constant}}{
                    store ($v_t$, $src_{stmt}$) in $DF$
                }
            }
        }
    }
    \;
    
    \SetKwProg{Fn}{Def}{:}{}
    \Fn{\FForwar{$m$, $stmt$, $var$}}{
        $tar_{stmt}$ = find all statements that contain $stmt$ \;
        \For{$s$ in $tar_{stmt}$}{
            store ($var$, $s$) in $DF$ \;
            \If{$s$ contain invoke statements}{
                $tar_{m}$ = method that invokes $stmt$ in $s$ \;
                $tar_{v}$ = variable in $tar_{m}$ that correspond to $var$\;
                ForwardAnalysis($tar_{method}$, $s$, $tar_{v}$)
            }
        }
    }

\end{algorithm}

\noindent\textbf{TPL DOI identification: }
We summarize PI from regulations~\cite{gdpr,ccpa,zhSDKLaw,piss} as listed in Tab.~\ref{tab:data_type}.
Then, we manually crawl PI-related APIs~\cite{androidDeveloper}.
Based on those APIs, {\oursystem} iterates over all statements and locates DOI with target variables. 
{\oursystem} performs inter and intra-procedural data flow analysis to identify all PI-related statements. 
For interested data types without official API, such as email and password, we use keywords matching method to identify potential data leakage.
%
Specifically, we user verb, i.e., ``get" and ``request" plus keywords, i.e., ``email" and ``contact", combination to mach those interested data. 

Algorithm~\ref{alg:data_usage} sketches the TPL data usage analysis method.
{\oursystem} iterates over all statements in TPL to locate the target variables ({\small{$var$}}) that represent personal information listed in Tab.~\ref{tab:data_type}.
Then, {\oursystem} conducts both inter-procedure and intra-procedure analysis to get the data flow.
For inter-procedures analysis, {\oursystem} backward analyzes statement {\small{$s_t$}} in method {\small{$m_t$}} with the DOI {\small{$v_t$}} (line 1).
The backward analysis locates the definition statements ({\small{$defs$}}) in {\small{$m_t$}} that relates to DOI {\small{$v_t$}} (line 11).
For statements in {\small{$defs$}}, {\oursystem} iterates {\small{$defs$}} and locates the variable signature of {\small{$v_t$}} in {\small{$d$}} (line 12-13).
{\oursystem} iterates FCG of target TPL and finds the methods ({\small{$caller_{list}$}}) that invoke {\small{$m$}} (line 14).
For each caller {\small{$m$}}, {\oursystem} locates the statements ({\small{$src_{stmt}$}}) in {\small{$m$}} that invoke {\small{$m_t$}}.
{\oursystem} finds the variable {\small{($src_v$)}} in {\small{$src_{stmt}$}} that corresponds to DOI {\small{$v_t$}} (line 17).
If {\small{$src_{v}$}} is a variable, {\oursystem} continues backward analysis, starts with {\small{$src_{stmt}$}} in {\small{$src_m$}} and focuses analysis on variable {\small{$src_v$}} (line 18-19); 
if {\small{$src_{v}$}} is a constant, {\oursystem} finds one piece of code that describes the data usage of {\small{$var$}} (line 21).

To locate the {\small{$v_t$}}'s data usage in {\small{$m_t$}}, {\oursystem} performs \textit{intra-method analysis}.
Specifically, {\oursystem} locates all statements ({\small{$uses$}}) that uses {\small{$var$}} in method {\small{$m$}} (line 5).
For each statement {\small{$u$}} in {\small{$uses$}}, {\oursystem} stores the usage of variable {\small{$var$}} (line 7) because the target variable has been used in the statement {\small{$u$}}.
Then, {\oursystem} conducts forward analysis (line 8) to identify whether {\small{$var$}} is used further.
%
%
For each statement {\small{$s$}} in {\small{$tar_{stmt}$}}, {\oursystem} stores the data flow describing the data usage. 
%
If {\small{$s$}} contains invoke statements, {\oursystem} locates the method {\small{$tar_m$}} that calls {\small{$stmt$}} (line 28), builds the variable {\small{$tar_v$}} corresponding to target variable {\small{$v_t$}} and continues the forward analysis (line 30).

\subsection{Identify data usage statements in TPL's privacy policy}
\label{sec:tpl_pp}

TPL privacy policy analysis aims to identify the data usage statements.
The pipeline of our privacy policy analysis consists of three steps: 
1) preprocessing of privacy policy,
2) named entity extraction, 
and 3) data usage action construction.
Finally, each privacy policy is abstracted into a set of \textbf{a}bstract \textbf{d}ata \textbf{u}sage \textbf{p}atterns (ADUP) that describes \textit{who} will (\textit{not}) \textit{access} \textit{what kinds of data} with \textit{whom}. 
Each ADUP is in form of {\small{\{data\_entity, action, \{data\_type\}, \{data\_recipient\}, \{neg\}\}}}.
For example, \textit{``We share your personal information with our service providers," (in Adad's privacy policy)} is abstracted as {\small{\textit{\{app, share, \{personal information\}, \{service provider\}, false\}}}}.

\subsubsection{Preprocessing privacy policy}
\label{sec:pre_pp}
To download the apps' privacy policies from Google Play~\cite{googleplay},
we use the \textit{Selenium Webdriver}~\cite{gojare2015analysis} to open websites of privacy policies.
Then, we use \textit{html2text}~\cite{html2text} to transform downloaded privacy policies into plain text. 
However, the transformed plain text contains many special characters, e.g. $*$ and $|$, which will lead to unnatural segmentation of sentences. 
{\oursystem} first uses patterns~\cite{stack_pre_nlp} to eliminate special characters. 
Besides, privacy policies use enumeration items~\cite{andow2019policylint} to detail related data items, which may lead to a whole sentence being segmented by item index and further cause losing information.
{\oursystem} merges the items by searching for sentences ending in a colon, and the next sentence starts with list items, such as bullets, roman numerals, and formatted numbers.

\subsubsection{Data access extraction}
With preprocessed privacy policy sentences, {\oursystem} analyzes each sentence to identify data access descriptions.
Since the interrogative sentences are commonly used in privacy policies but do not describe the types of accessed personal information, we omit them by checking the interrogative words (i.e., who, where, when, why, whether, what, and how). 
For example, the sentence ``\textit{\underline{What} personal information do we collect"} is omitted. 
Then, we construct the part-of-speech (POS) tagging, and dependency parsing tree (DPS) for each sentence.
We will omit sentences without verb words because those sentences will not claim data access behaviors.
For example, ``\textit{Do-Not-Track Signals and Similar Mechanisms}" in \textit{Facebook Open Source Privacy Policy} only claims the title of the paragraph.
For the remaining sentences, we follow the following steps to extract data action, data entity, and data actor.

%
\noindent\textbf{Identification of data action words}. 
We first identify whether the verb words are in the data usage word list (Tab.~\ref{tab:actionwords}).
We stem the word using \textit{nltk}~\cite{loper2002nltk} to deal with English verb tenses. 
We also identify the modifier of the target verb because the privacy policies also claim the data they will not use or collect.
To achieve this, we analyze the dependency parsing tree and recognize the words whose relation with the target verb is ``\textit{advmod}" (adverbial modifier) or ``\textit{mmod}" (modal verb modifier).
If the words or their tags are negative (i.e., ``neg"), we regard the data usage action as negative one.
To this end, {\oursystem} identifies data sharing or collection operations.

\begin{table}[]
\centering
\smaller
\vspace{2ex}
\caption{Personal information and corresponding APIs.}
\scalebox{0.95}{
\begin{tabular}{@{}p{0.07\textwidth}p{0.23\textwidth}p{0.14\textwidth}@{}}
\toprule
\textbf{PI} & \textbf{Class Name} & \textbf{Method Name} \\ \midrule
Ad ID & AdvertisingIdClient & getadvertisingidInfo \\
AD ID &  AdvertisingIdClient & getInfo \\
\rowcolor[HTML]{EFEFEF} 
Bluetooch address & android.bluetooth.BluetoothAdapter & getAddress \\
Camera                         & android.hardware.Camera                                  & setPreviewDisplay                            \\
\rowcolor[HTML]{EFEFEF} 
Cell   Loca-tion                & android.telephony.TelephonyManager                       & getCellLocation                              \\
Contact                        & android.provider.ContactsContract                        & PhoneLookup                                  \\
                               & \cellcolor[HTML]{EFEFEF}android.location.Location        & \cellcolor[HTML]{EFEFEF}getLatitude          \\
                               & android.location.Location                                & getLongitude                                 \\
                               & \cellcolor[HTML]{EFEFEF}android.location.LocationManager & \cellcolor[HTML]{EFEFEF}getLastKnownLocation \\
                               & android.location.LocationManager                         & requestLocationUpdates                       \\
\multirow{-5}{*}{GPS Location} & \cellcolor[HTML]{EFEFEF}android.location.LocationManager & \cellcolor[HTML]{EFEFEF}getLongitude         \\
Sim Number                     & android.telephony.TelephonyManager                       & getSimSerialNumber                           \\
\rowcolor[HTML]{EFEFEF} 
IMEI                           & android.telephony.TelephonyManager                       & getDeviceId                                  \\
IMSE                           & android.telephony.TelephonyManager                       & getSubscriberId                              \\
\rowcolor[HTML]{EFEFEF} 
TimeZone                       & java.util.Calendar                                       & getTimeZone                                  \\
MAC address                    & android.net.wifi.WifiInfo                                & getMacAddress                                \\
\rowcolor[HTML]{EFEFEF} 
Password                       & android.accounts.AccountManager                          & getPassword                                  \\
Phone   Number                 & android.telephony.TelephonyManager                       & getLine1Number                               \\
\rowcolor[HTML]{EFEFEF} 
SMS                            & android.telephony.SmsManager                             & sendTextMessage                              \\
SSID                           & android.net.wifi.WifiInfo                                & getSSID                                      \\
\rowcolor[HTML]{EFEFEF} 
User Credential                & android.accounts.AccountManager                          & getAccounts                                  \\
UserName                       & android.os.UserManager                                   & getUserName                                  \\ \bottomrule
\end{tabular}
}
\label{tab:pi_api}
\end{table}

\noindent\textbf{Identification of data entity}.
We identify objects of identified collecting and sharing words (target verb) according to DPS.
{\oursystem} first merges the noun phrase (NP) tag into one noun tag according to POS tagging to simplify sentence structure. 
%
For example, consider the following sentence in the privacy policy of the app \textit{video.reface.app} ``\textit{\small{With the use of Google Analytics, we collect such data as IP-Address, your \underline{device model}, \underline{screen resolution} and \underline{operation system}, \underline{session duration}, your location}}''.
%
%
As only one noun or noun phrase will be identified as the object of the target verb~\cite{hanlp21}, {\oursystem} locates all coordinating nouns for the object. 
Specifically, {\oursystem} iterates the constituency parsing tree, which is calculated by two stages conditional random field (CRF) model~\cite{zhang2021fast}, to find the sub-trees whose \textit{a) root label are common noun (NN tag) or Noun Phrase (NP)} or \textit{b) coordinating conjunction (CC tag) or punctuation exist in the sub-tree}.
The condition \textit{a)} denotes that the leaves in the tree can be merged into one noun, and  \textit{b)} indicates there may exist enumeration items or coordinating nouns.
If we omit the relations, {\oursystem} cannot fully extract all data entities, i.e., false-positive occurs.
To this end, we obtain all candidate data objects in each sentence.

\noindent\textbf{Identification of data actor.}
To identify the data actor, {\oursystem} analyzes DPT and locates the words whose relation with sharing and collection verbs is nominal subject (``\textit{nsubj}") or direct object (``\textit{dobj}").
Suppose the sentence or the object is a prepositional complement or clausal complement of a preposition (``\textit{ccomp}").
We iterate the DPT to find the coordinating verb and then locate the nominal subject of the target verb.
The nominal subject and the verb will be added to the actor list and the sharing/collection verb list, respectively.
We also consider the data actor as users separately and the situation when the real data actor is the user rather than the supplier.
For example, ``\textit{we will not collect the personal information you shared with us}''.
The data actor should be ``you"  while the data recipient is ``us".

\noindent\textbf{Post-process}
To eliminate false detection, e.g., conditional clauses, we design rules to avoid false data usage extractions.
For sentences that start with condition or assumption hints, such as \textit{`if'} (\textit{``if you do not provide your personal information"}), we will not identify the data usage pattern inside.

\begin{table}[b]
\centering
\smaller
\caption{Data sharing and collecting words.}
\begin{tabular}{p{0.23\columnwidth}p{0.67\columnwidth}}
\toprule
\textbf{Action Type} & \textbf{Keywords} \\ \midrule
Collect &  access, check, collect, gather, know, obtain, receive, save, store, use \\ \hline
Sharing &  accumulate, afford, aggregate, associate,  cache,  combine, convert, connect,  deliver, disclose, distribute, disseminate, exchange,  gather, get, give, keep, lease, obtain, offer, post, possess, proxy,  provide, protect against, receive,  rent, report, request, save, seek, sell, share, send, track,  trade, transport, transfer, transmit\\ \bottomrule
\end{tabular}
\label{tab:actionwords}
\end{table}

\subsection{Extract host apps' interaction with TPLs}
\label{sec:host_app_data}

To check whether host apps' privacy policies are consistent with their TPL usage, 
{\oursystem} conducts static analysis to identify the host apps' interaction with TPLs.
{\oursystem} discloses what kinds of PI data (Tab.~\ref{tab:data_type}) are shared with TPLs.
{\oursystem} uses flowdroid~\cite{arzt2014flowdroid} to construct FCG of apps.
Since flowdroid optimizes Android app analysis process~\cite{arzt2014flowdroid}, {\oursystem} directly uses the FCG constructed from flowdroid.
{\oursystem} performs the data flow analysis (\S\ref{sec:tpl_data}) to find the PI related data flow.
To identify TPLs, {\oursystem} identifies whether the statement contains the TPLs' name or package name or not.
After obtaining TPLs and PI data, we identify whether the method exists in the data flows.
Besides, we optimize the signature of method invocation (\textit{SMI}) using the invoking statement.
{\oursystem} matches the \textit{SMI} in both PIs' data flow and TPLs' data flow.
Once the \textit{SMI} is matched between $TPL_i$ and $PI_j$, {\oursystem} regards the host app sharing the $PI_j$ with the $TPL_i$.
If the \textit{SMI} only exits in $PI_j$'s data flow, {\oursystem} regards that the host app collects $PI_j$  without sharing it with TPLs.

\subsection{Identify TPLs usage statements in host apps' privacy policy}
\label{sec:host_app_pp}

To identify the TPL usage statements in host {\app}s, we extract two descriptions from the privacy policy: 
a) the \textit{data sharing} related statements and 
b) TPL-related descriptions.
The \textit{data sharing} related statements claim how the {\app} discloses users' personal information under which situations.
For example, the {\app} (\textit{flipboard.boxer.app}) claims that "\textit{We share personal information with vendors and service providers that help us offer and improve our service}".
%
The statement indicates that \textit{flipboard.boxer.app} will share users' \textit{personal information} with vendors and service providers.
If {\oursystem} also identifies the {\app}'s data flow sharing users' data with TPLs tagged as vendors or service providers, {\oursystem} regards the app's privacy policy as consistent with its behavior.

After preprocessing privacy policies (\S\ref{sec:pre_pp}), we use NLP methods~\cite{hanlp21} to get the tokenization, POS tagging and DPS for each sentence.
For sentences with verbs, we check whether the verb is in \textit{sharing} word list (Tab.~\ref{tab:actionwords}).
After obtaining the data sharing verb, we identify the data recipient to determine which TPL the data are shared with.
Specifically, we iterate the words in the sentence.
If the word is the object of a preposition (\textit{pobj}) or an indirect object (\textit{iobj}) of the verb, we identify the words as candidate data recipients, i.e., TPLs.
For enumeration patterns, once we identify the sentence that a) starts with a noun or noun phrase (denoted as $tar_{TPL}$) that is involved in our TPL list, v) the noun ends with colon following only noun or noun phrases, we regard the noun (phrases) following colon as the shared data with $tar_{TPL}$.

\subsection{TPL privacy compliance investigation}

{\oursystem} investigates TPL privacy compliance by performing 
\textit{normativeness analysis}, 
\textit{privacy policies legality analysis}, 
and \textit{privacy non-compliance transmissibility analysis}.

\noindent\textbf{Normativeness analysis} identifies whether TPL provides privacy policies.
According to GDPR Article 12(1) and Article 13(1), TPLs, who act as the data controller, or third party of personal data, should infer users about data access behaviors in a concise, transparent, intelligible and easily accessible form.

\noindent\textbf{Legality analysis} identifies conflicts between TPLs' behavior and privacy policy statements.
Specifically, {\oursystem} identifies whether TPLs' data access actions are clearly claimed in their privacy policies by combining the data flow results (\S\ref{sec:tpl_data}) and PP ADUP (\S\ref{sec:tpl_pp}).
If {\oursystem} identifies data usage in TPL data flow but the results are not mentioned in \textit{$data\_type$}s of ADUP set, the TPL will be regarded as violating legal requirements.

\noindent\textbf{Privacy non-compliance transmissibility analysis} 
\kaifa{checks to what extent the TPLs (TPL$_v$), whose privacy policies violate the regulation requirements, affect other TPLs or apps.
{\oursystem} summarizes artifacts that use TPL$_v$ from {\textit{Maven Repository}} to investigate the impact of TPL's privacy non-compliance.
Specifically, {\oursystem} crawls \textit{Usages} information in {\textit{Maven Repository}} to count the number of artifacts that use TPL$_v$.
}

\subsection{Regulation issue analysis}

Regulation conflict analysis discovers whether the usage of TPLs in host apps comply regulation requirements.
Two kinds of conflict will be identified:

\noindent{$\bullet$} {\oursystem} identifies whether usage of TPLs from data flow analysis (\S\ref{sec:host_app_data}) is clearly claimed in their privacy policy statements (\S\ref{sec:host_app_pp}). 
{\oursystem} investigates whether the TPL package names in data flow analysis match the \textit{$data\_entity$} or \textit{$data\_recipient$} in ADUP of host app privacy policy results.
    
\noindent{$\bullet$} {\oursystem} discloses whether the host apps clearly state their data interaction with TPLs. 
{\oursystem} identifies whether the apps' TPL interaction behavior (\S\ref{sec:host_app_data}) is clearly stated in host apps' privacy policies (\S\ref{sec:host_app_pp}).

\section{Evaluation}
\label{sec:eva}
\label{sec:dataset}

We evaluate the performance of {\oursystem} by answering the following research questions:

\noindent\textbf{RQ1: Normativeness Analysis of TPLs.} 
\textit{How many TPLs provide privacy policy documents?} 

\noindent\textbf{RQ2: Legality Analysis of TPLs.}  
\textit{Do TPLs’ privacy policies meet regulation requirements, and do TPLs’ privacy policies correctly claim data usage actions?}

\noindent\textbf{RQ3: Host {\app}s behavior analysis.} \textit{Do host {\app}s conduct privacy data interaction with TPLs?}

\noindent\textbf{RQ4: Legality of host app's privacy policies.} 
\textit{Do host app's privacy policies comply with  requirements for disclosure of TPL usage?}


\begin{table}[t]
\centering
\vspace{3ex}
\smaller

\caption{List of TPLs without Privacy Policies.}
\scalebox{0.95}{%
\begin{tabular}{@{}p{0.12\textwidth}p{0.24\textwidth}p{0.1\textwidth}@{}}
\toprule
\textbf{TPL Categories}         & \textbf{TPL Name}               & \textbf{Reasons} \\ \midrule
\multirow{2}{*}{Ad Network}     & Fractional Media                & NOS              \\
                                & YuMe                            & NOS              \\ \midrule
\multirow{2}{*}{Social library} & Smack API                       & NPOS             \\
                                & Twitter4j                       & NPOS             \\ \midrule
Development Tool                & Android In-App Billing Library  & GNP              \\
                                & Apache Commons Codec            & NPOS             \\
                                & Apache Commons I/O              & NPOS             \\
                                & Apache Commons Lang             & NPOS             \\
                                & Apache Commons Logging          & NPOS             \\
                                & Apache Http Auth                & NOS              \\
                                & Apache HttpMime API             & NOS              \\
                                & Apache James Mime4j             & NPOS             \\
                                & Apache Thrift                   & NPOS             \\
                                & AChartEngine                    & GNP              \\
                                & FasterXML Jackson               & GNP              \\
                                & Android ViewBadger              & GNP              \\
                                & Kin                              & GNP              \\
                                & Material App Rating             & GNP              \\
                                & OpenStreetMap tools for Android & GNP              \\
                                & Android GIF Drawabl             & GNP              \\
                                & Crouton                         & GNP              \\
                                & Google GData client             & GNP              \\
                                & Google Guava                    & GNP              \\
                                & Google gson                     & GNP              \\
                                & HttpClient for Android          & GNP              \\
                                & JSON.simple                     & GNP              \\
                                & greenDAO                        & NPOS             \\
                                & libgdx                          & NPOS             \\
                                & Material DateTime Picker        & GNP              \\ \midrule
\multicolumn{3}{p{0.45\textwidth}}{\small{NOS: No Official webSite; NPOS: No Privacy policies  on Official webSite; GNP: Github project without Privacy policies}}
\end{tabular}%
}
\vspace{-3ex}
\label{tab:tpl_no_pp}
\end{table}

\subsection{RQ1: Normativeness Analysis of TPLs.}
\label{sec:rq1}

\noindent\textbf{Experiment Setup.}
This research question evaluates all Android TPLs listed in AppBrain~\cite{appbrain} including three types: Ad networks, social libraries, and development tools.
The list gives the TPLs that are widely used by top 500 installed apps in Google Play. 
The data set contains 458 TPLs that include 141 ad networks, 25 social libraries, and 292 development tools.
Then, we gather the privacy policies of those TPLs by visiting the homepage provided by each TPL information page on AppBrain.
We manually crawl information of TPLs whose homepages are not provided. 
For TPLs in Github or Bitbucket, we visit their public repositories and check whether privacy policies are provided.
When visiting the homepage of TPLs, we search the homepage with keywords, such as \textit{privacy, policy, legal} and \textit{policies}, to find potential privacy policy links on the homepage. 
Besides, we also manually check whether policy links are given in the typical layout ~\cite{harkous2018polisis}, such as the bottom of the website.
%

\noindent\textbf{Results.}
According to regulation requirements (\S\ref{sec:bb_regulation_req}), data controllers should clearly disclose their data usage in privacy policies.
This research question discloses whether TPLs satisfy the requirements of regulations, i.e., do TPLs correctly provide privacy policies.
With the TPL list and  privacy policies, 
we discover that 21 of 141 Ad network TPLs, 10 of 25 social libraries, and 180 of 292 development tools TPLs, which account for 31\%, do not provide privacy policy websites. 
We find that some TPLs are created by individual developers and released on {\github}~\cite{github}.
We manually crawl their privacy policies for {\github} repositories.
Tab.~\ref{tab:tpl_no_pp} summarizes partial TPL names and the reasons that the TPLs do not provide privacy policies.
Tab.~\ref{tab:tpl_no_pp} demonstrates that 16 of 22 TPLs that do not provide privacy policies are published as Github repositories (GNP).
Five TPLs without privacy policies miss official websites (NOS), and the left does not provide privacy policy documents on their websites (NPOS).
For development tool TPLs, we also observe that 66 TPLs from \textit{Google Inc.} share the same privacy policy~\cite{googlepp} that may lead to over-claiming the personal information usage for specific TPLs.
For nine TPLs from \textit{Apache}, six TPLs (6/9) do not provide privacy policies and two TPLs' (2/9) websites are Not Found.

\answer{1}{
{\oursystem} reveals that 31\% TPLs do not provide privacy policies. Over 14\% TPLs from the same company provide one general privacy policy.
}


\begin{figure}[t]
    \centering
    \vspace{-1ex}
    \includegraphics[width=0.4\textwidth]{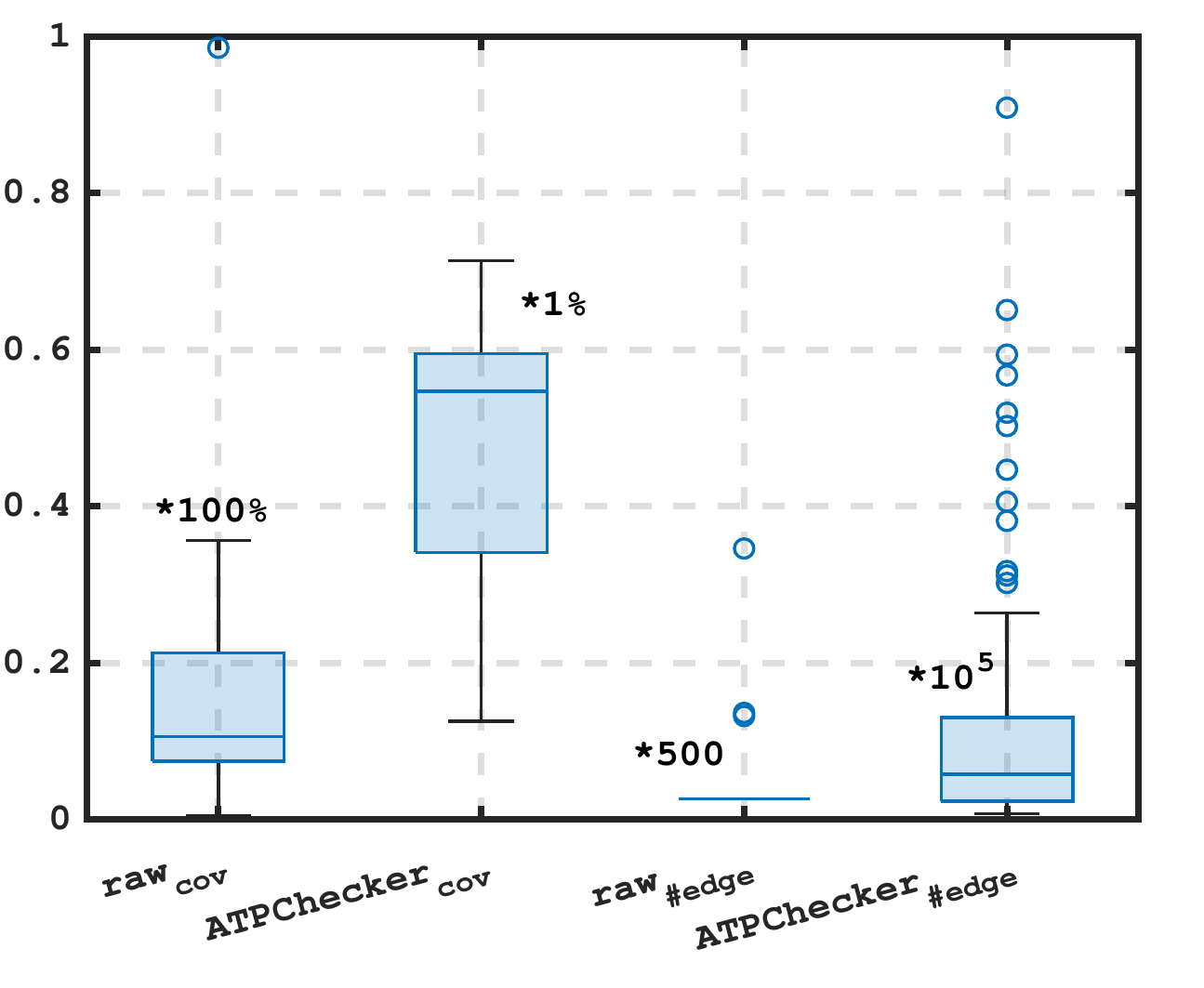}
    \vspace{-1.5ex}
    \caption{Evaluation of function call graph construction.}
    \label{fig:fcg_compare}
\end{figure}

\subsection{RQ2: Legality Analysis of TPLs.}

\noindent\textbf{Experiment Setup.}
We collect the resources of TPLs from Maven Repository~\cite{mvnrepository} based on the TPL list from AppBrain (\S\ref{sec:rq1}).
We manually crawl TPL binary files from Maven Repository by searching the TPL's name in Appbrain's list and finally get 187 different TPLs including 87 ``aar" and 100 ``jar" files.
More specifically, we get 45 ad networks, 132 development tool libraries and 10 social libraries.

\noindent\textbf{Results.}
{\oursystem} discloses whether the TPLs' privacy policies satisfy the regulation requirements by comparing TPLs' binary files analysis results (\S\ref{sec:tpl_data}) and privacy policies analysis results (\S~\ref{sec:tpl_pp}).
{\oursystem} analyzes TPLs' privacy policies and obtains the ADUP of personal information usage-related statements.
{\oursystem} identifies personal information in the TPL through static analysis (\S\ref{sec:tpl_data}).
Finally, {\oursystem} matches the consistency between AUDP and personal information.



\begin{figure}[t]
    \centering
    \includegraphics[width=0.4\textwidth]{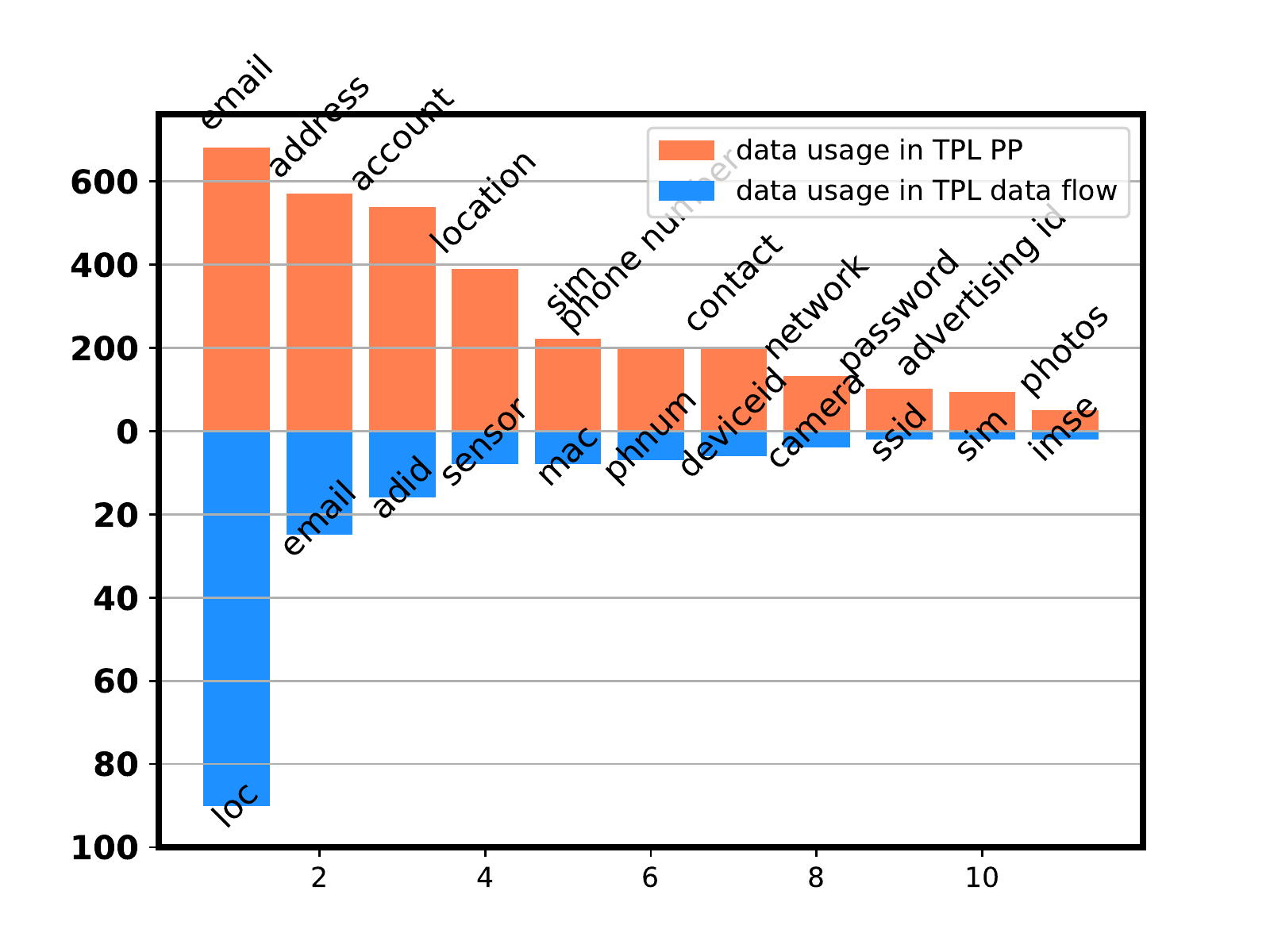}

    \caption{Popularity of data types that are used in TPLs' data flow and PP.}
    \vspace{-1ex}
    \label{fig:tpl_data_freq}
\end{figure}

\noindent{\small{\textit{\textbf{TPL privacy policies analysis.}}}}
{\oursystem} analyzes TPLs' privacy policy documents.
We analyze 52,523 sentences with sharing and collection (SAC) words.
Among sentences with SAC words, 15,270 sentences start with 5W1H (who, why, when, whether, what, how) words that are considered as data usage actions.
Besides, 5,205 sentences only vaguely state that the TPL will collect personal information without specific data types.
For example, $com.xiledsystems$  only claims \textit{``We collect your personal information in order to provide and continually improve our products and services,"} but no specific personal information is given.
In data usage-related actions, {\oursystem} analyzes the personal information that is mostly used by TPLs.
Fig.~\ref{fig:tpl_data_freq} shows the times that are mentioned in TPLs' privacy policies.
It can be observed that \texttt{contact} is the most  used personal information.
This may be  caused by the fact that, by getting contact, the TPLs can easily expand and prompt their services.
Account, address and email are the second most popular data mentioned in TPLs' privacy policies.
Among the statements that claim to collect  accounts, addresses, and email, 21.13\%  sentences claim that they collect related data for contacting users. 
For example, one of them claims that \textit{``we will use your email, phone number, or other contact information you provide us by written or oral means for contacting you and providing you with the services and information that             you request"}.

\vspace{1ex}
\noindent{\small{\textit{\textbf{TPL data usage analysis.}}}}
We first evaluate the effectiveness of {\oursystem} in improving the FCG construction performance.
We analyze TPLs' FCG quality through comparing the a) FCG coverage and b) the number of edges between the raw FCG built by soot and our optimized FCG.
FCG coverage is calculated by:
\begin{equation}
    \label{eq:fcg_cov}
    cov_{FCG} = \frac{\#Nodes_{FCG}}{\#Nodes_{TPL}},
\end{equation}
where {\smaller{$\#Nodes_{FCG}$}} is the number of nodes (methods) in FCG and {\smaller{$\#Nodes_{TPL}$}} is the total number of TPLs' methods.
Fig.~\ref{fig:fcg_compare} illustrates the box plots of {\oursystem}'s performance in  optimizing  FCG construction, and the number at the top of boxes denotes measuring units of the data for better illustration.
Fig.~\ref{fig:fcg_compare} demonstrates that {\oursystem}'s FCG coverage achieves an average performance of 55\%, while the FCG coverage of raw FCG only achieves an average of 0.34\%, which  means it nearly fails to construct the FCG of TPLs.
This could be caused by the fact that soot can not identify comprehensive entry points of TPLs to construct the FCG and thus cause many nodes and edges are missed.
{\smaller{$raw_{\#edge}$}} denotes the raw FCG only contains the 159 average edges, while the {\smaller{${\oursystem}_{\#edge}$}} shows the improved FCG can extract 12,960 edges.
The results indicate that the soot nearly fails to analyze the FCG of TPLs because of the misidentification of TPL's entry points for FCG construction, and {\oursystem} optimizes construction performance and improves the FCG quality.

{\oursystem} analyzes the data usage of TPLs to identify the consistency with their privacy policy statements.
{\oursystem} analyzes the data flow of TPLs and discovers that 38 out of 187 TPLs access users' personal information.
{\oursystem} identifies 176 data access traces.
Among data usage behaviors, 90 out of 176 traces access  location information.
Fig.~\ref{fig:tpl_data_freq} shows the data frequencies that are mostly used in TPLs' data flow (shown in orange color).
It can also be observed that TPLs' privacy policies tend to disclose non-identical information while their actions tend to access identical information, such as location.

\begin{figure}[t]
    \centering
    \includegraphics[width=0.4\textwidth]{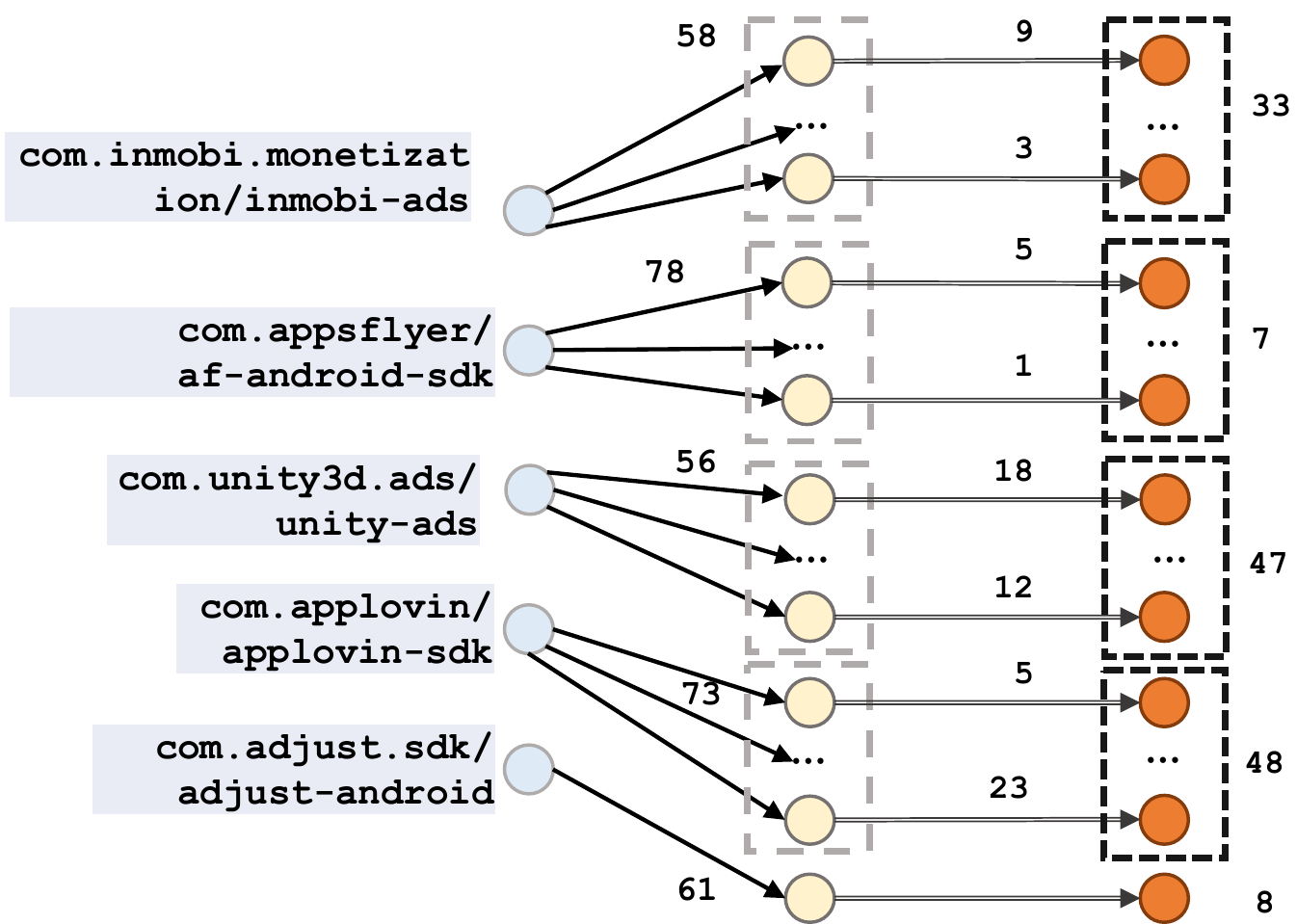}

    \caption{TPL Privacy Policy Non-compliance Propagation. The first column is raw TPLs identified by ATPChecker that potentially violate regulation requirements. The second column is TPLs affected by raw TPLs and the number above the arrow is the number of TPLs infected by raw TPLs. The third column is TPLs affected by the TPLs in the second column.}
    \label{fig:tpl_pp_dep}

\end{figure}

{\oursystem} checks the compliance of TPLs by combing the results of privacy policies analysis and data flow analysis.
Specifically, {\oursystem} identifies how many TPLs comply with regulations by checking whether the PI in data flow is also mentioned in ADUP. 
%
{\oursystem} identifies 38 TPLs access users' data.
Among those 38 TPLs, {\oursystem} discovers that 15 out of 38 (39.5\%) TPLs violate the regulation requirements, i.e., the TPL collects at least one users' data without clearly disclosing them in the TPL's privacy policy.

\noindent{\small{\textit{\textbf{Investigation on perniciousness of TPLs' non-compliant behavior.}}}}
After identifying TPLs that do not satisfy the regulation requirements, we also study the effect of TPLs' non-compliant behavior on a large scale.
%
TPLs may integrate other TPLs to enhance their functionalities or facilitate usability~\cite{atvhunter2021ICSE}.
Thus, once one TPL ($TPL_v$) violates the regulation, other artifacts using $TPL_v$ may spread the threats and make their privacy policies violate regulation requirements.
To expose the impacts, {\oursystem} analyzes the dependencies among TPLs. 
Specifically, {\oursystem} crawls the TPLs ($TPL_U$) that use $TPL_v$, which are analyzed in \S\ref{sec:tpl_data}-\ref{sec:tpl_pp}, from {\maven}.
Then starting from $TPL_U$, {\oursystem} crawls the list of artifacts that use $TPL_U$.
In this way, {\oursystem} investigates the propagation of $TPL_v$'s threats under two times integration and visualizes it in Fig.~\ref{fig:tpl_pp_dep}.
{\oursystem} starts the analysis with 15 unconventional TPLs.
Fig.~\ref{fig:tpl_pp_dep} shows the effect of five TPLs whose privacy policies have inconsistency issues.
%
It can be observed that after one round of propagation, even five TPLs will affect 321 TPLs (15 TPLs affects 434 TPLs). After two rounds of propagation, the threats even spread to extra 143 TPLs (15 to 168).
%
Fig.~\ref{fig:tpl_pp_dep} also illustrates that both popular and minority TPLs can have a significant impact on the propagation of privacy non-compliance. 
Popular TPLs are widely used by other TPLs, making them highly infective, while minority TPLs can also affect a large number of TPLs.
After two rounds of propagation, the number of TPLs they affect increases exponentially.
This observation indicates that developers should pay attention to the usage data of TPLs, especially the functions that are related to privacy.

\answer{2}{
{\oursystem} identifies that over 39.5\% TPLs miss disclosing their data usage in privacy policy documents.
{\oursystem} investigates that the effect of the privacy policies with non-compliance issues spreads widely.
}

\subsection{RQ3: Analysis of host apps' interaction with TPLs}
\label{sec:rq3}

\noindent\textbf{Experiment Setup.}
We collect the host apps of TPLs from AppBrain to investigate host apps' interaction with TPLs.
We gather the host app list using the mapping relations between host apps and used in-app TPLs provided in AppBrain.
%
%
For TPL's host apps, we can only access 10 apps which are mostly downloaded in Google Play.
Finally, we gather a total of 641 distinct apps because some apps may use multiple TPLs and one TPL can be used by different apps.
%

\begin{figure}[t]
    \centering
    \includegraphics[width=0.45\textwidth]{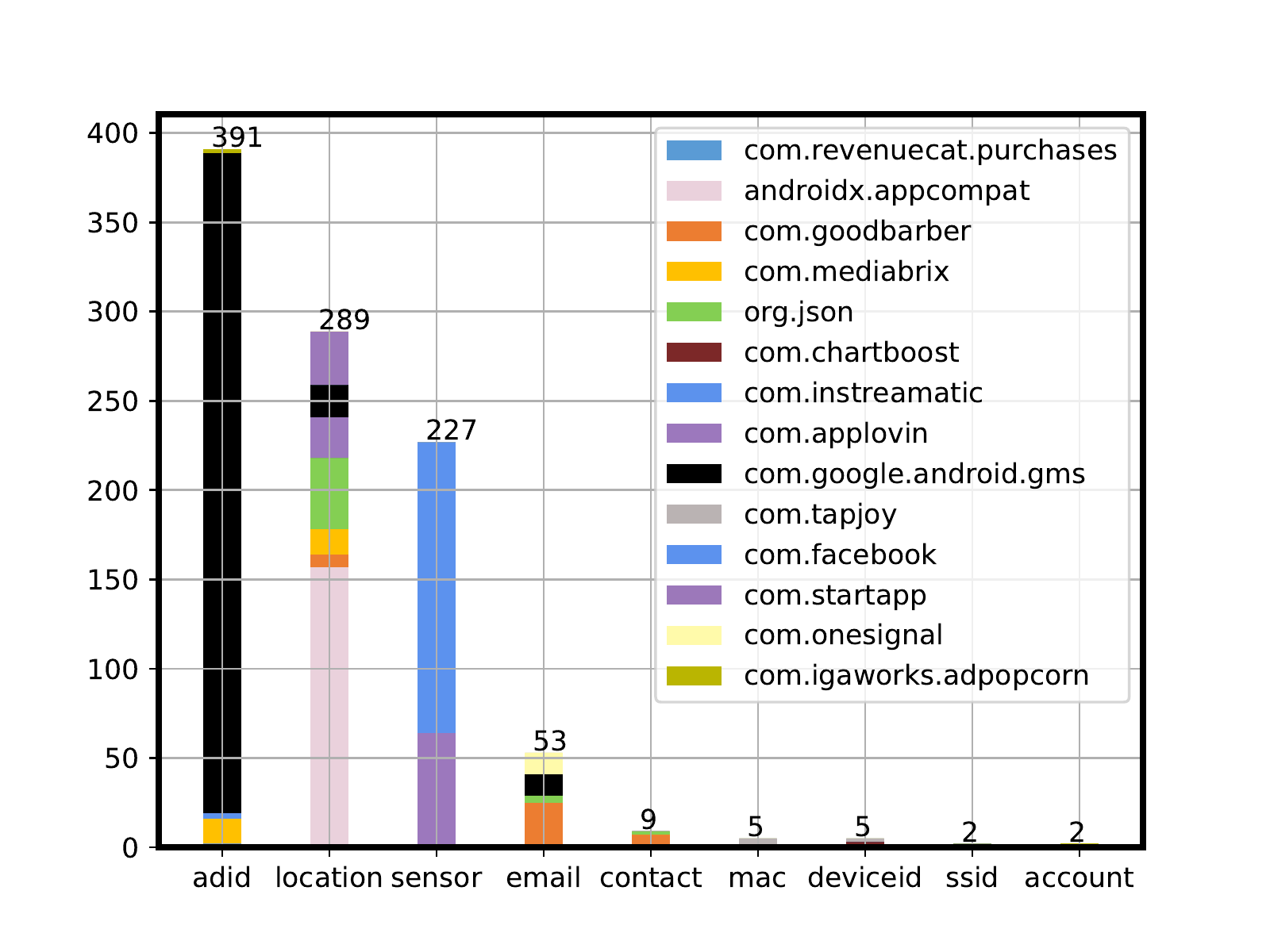}
    \caption{Statistics of host apps' data interaction with TPLs.}
    \label{fig:app_int_tpl}
\end{figure}

\noindent\textbf{Results.}
{\oursystem} performs data flow analysis to investigate whether host apps conduct personal information interaction with TPLs and what kinds of PI are shared with TPLs.
{\oursystem} analyzes 641 distinct apps. 
Due to the restraint of computational resources and the limitations of flowdroid, we successfully analyzed 459 apps.  174 out of 641 apps are in the form of ".xapk"~\cite{apkpure} which cannot be analyzed by flowdroid, and the others cannot be analyzed because some apps apply protection methods and some apps are too large which exhaust our computation resources~\cite{qiu2018analyzing}.
{\oursystem} identifies that over 47.9\% (220/459) host apps share PI with TPLs.
%
{\oursystem} measures the suspicious behavior by counting the times of PI-related invocations in apps' data flow and traces 973 times data-sharing behaviors with TPLs and finds that over 40.2\% traces share users' \textit{advertising ID} with TPLs such as \textit{``com.google.android.gms"}.
%
%
Fig.~\ref{fig:app_int_tpl} shows the nine types of data shared by host apps with TPLs.
Fig.~\ref{fig:app_int_tpl} shows that the second popular data the host app need to share with TPLs is location.
{\oursystem} also observes that \textit{``com.google.android.gms"} is the TPL that accesses the most PI from host apps.
Although \textit{``com.google.android.gms"} provides abundant functionalities, such as Gmail and music, TPLs should avoid collecting information that can be used to easily identify an individual and provide privacy policies to state the purpose of data access.
TPLs should not ask for data which is unnecessary for their services~\cite{zhSDKLaw,gdpr,ccpa,piss}.

\answer{3}{
{\oursystem} identifies that over 47.9\% of host apps share personal information with TPLs.
Among shared PI, advertising ID accounts for over 40.2\%  of traces in identified user data access actions.
{\oursystem} also discovers that while well-known TPLs provide rich functionalities, they also request personal information.
}

\subsection{RQ4: Analysis of host apps' privacy policy.}
\label{exp:host_app_pp}

\noindent\textbf{Experiment Setup.}
We collect the privacy policies of host apps to determine whether the apps' data usage behavior is consistent with the privacy policies.
For each host app in our dataset, we  get its privacy policy by crawling its homepage from Google Play.
To evaluate the performance of ATPChecker, we collect extra 254 apps, which can be successfully analyzed, and their privacy policies.  
To annotate these privacy policies, one expert and senior researcher, who have conducted research on privacy policy and software engineering for over seven years and two years respectively, manually label the sentences that include data usage, TPL usage, and data interaction with TPLs.
The annotation process is conducted in an in-house manner~\cite{sorokin2008utility,zhao22ca4p} that guarantees the quality and agreement of the labels.

\noindent\textbf{Metrics.} 
We conducted an evaluation of ATPChecker's performance in identifying the accuracy of host apps' claims regarding i) the usage of TPLs in their privacy policies, and ii) TPLs' data usage in host apps.
Considering ATPChecker's purpose in identifying compliance, we define metrics as follows.
For i), which focuses on whether ATPChecker can correctly identifies the host apps' usage and declaration of TPL usage in their privacy policies:

\noindent\textit{\small{\textbf{True Positive}}}: The host app uses the TPL without clarifying it in the host apps' privacy policies (non-compliance). ATPChecker identifies the host app’s usage of TPL and cannot identify the claim of the TPL’s usage in the host app’s privacy policy. 
It is worthy noticing that if ATPChecker cannot identify the usage of TPL in host app’s Code, ATPChecker cannot identify the positive behavior, i.e., the host app uses the TPL without claiming the usage of TPL in app’s privacy policy.

\noindent\textit{\small{\textbf{True Negative}}}: The host app declares the usage of TPLs in their privacy policies (compliance).  
i) ATPChecker identifies the host app’s TPL usage in the host app’s binary files and identify the claim of TPL usage in the host app’s privacy policy. 
ii) ATPChecker identifies the claim of TPL usage in the host app’s privacy policy.

\noindent\textit{\small{\textbf{False Positive}}}: The host app declares the usage of TPLs in their privacy policies (compliance). 
i) ATPChecker identifies the host app’s usage of TPL and cannot identify the claim of the TPL’s usage in the host app’s privacy policy. 
ii) ATPChecker cannot identify the usage of TPL in app’s code and cannot identify the claim of TPL usage in app’s privacy policy.

\noindent\textit{\small{\textbf{False Negative}}}: The host app does not claim the usage of TPL in the app's privacy policy.
i) ATPChecker identifies TPL usage in host app’s binary files and also identify the claim of TPL usage in the app’s privacy policy.
ii) ATPChecker identifies the claim of TPL usage in the app's privacy policy.

For the second aspect (ii), which assesses whether ATPChecker can correctly identify the host apps' declaration of TPLs' data usage in the app's privacy policies:

\noindent\textit{\small{\textbf{True Positive}}}:
The host app shares user data with TPL without clarifying it in the host app’s privacy policies (non-compliance). ATPChecker identifies the host app’s data-sharing behavior in host app’s code and cannot identify the claim of data-sharing behavior in the app’s privacy policy.
Please note that if ATPChecker cannot identify the host app’s data-sharing behavior with TPL in its Code, ATPChecker cannot identify the positive behavior, i.e., the host app shares user data with TPL without clarifying it in the host app’s privacy policies.

\noindent\textit{\small{\textbf{True Negative}}}:
The host app shares user data with TPL and clarifies it in the host app’s privacy policies (compliance). ATPChecker identifies the host app’s data-sharing behavior and identifies the claim of the data-sharing behavior in the app’s privacy policy.
Please note that if ATPChecker cannot identify the host app’s data-sharing behavior with TPL in its Code, ATPChecker cannot identify the positive behavior, i.e., the host app shares user data with TPL and clarifies it in the host app’s privacy policies.

\noindent\textit{\small{\textbf{False Positive}}}:
The host app demonstrates compliance by sharing user data with TPLs and explicitly stating this behavior in the host app's privacy policies.
i) ATPChecker successfully identifies the host app's data-sharing behavior within the host app's code; however, it is unable to identify the explicit claim of data-sharing behavior in the app's privacy policy.
ii) ATPChecker does not have the capability to identify the app's data-sharing behavior as described in the app's privacy policy.
Furthermore, it is important to note that due to the unavailability of ground truth regarding the host apps' data-sharing behavior and their claims in the privacy policies, the assessment of false positives by ATPChecker cannot be evaluated in the current version.


\noindent\textit{\small{\textbf{False Negative}}}:
The host app shares user data with TPL without clarifying it in the host app’s privacy policies or the app does not claims the data sharing with TPLs. 
i) ATPChecker identifies the host app’s data-sharing behavior and identifies the claim of the data-sharing behavior in the app’s privacy policy.
ii) ATPChecker only identifies the the claim of the data-sharing behavior in the app’s privacy policy

\begin{table}[t]
\vspace{3.5ex}
\centering
\smaller

\caption{Evaluation of {\oursystem} for identifying host apps' compliance. }
\smaller
\label{tab:metrics}
\begin{tabular}{@{}c|cccc|cccc@{}}
\toprule
& \multicolumn{4}{c|}{TPL List} & \multicolumn{4}{c}{TPL Data}           \\ \midrule
\multirow{4}{*}{{\begin{tabular}[c]{@{}c@{}}Trace \\ Level\end{tabular}}} & TP & TN & FP & FN & TP & TN & FP & FN \\
 & 87 & 219 & 0 & 22 & 143 & 123 & / & 39 \\ \cmidrule(l){2-9} 
 & A   & P & R & F1 & A & P & R     & F1 \\
 & 93.29\% & 1 & 79.82\% & 88.78\%  & /        & /         & 78.57\% & /  \\ \midrule
\multirow{4}{*}{\begin{tabular}[c]{@{}c@{}}App \\ Level\end{tabular}}   & TP & TN & FP & FN & TP & TN & FP & FN \\
 & 56 & 118 & 0 & 24 & 62 & 57 & / & 22 \\ \cmidrule(l){2-9} 
 & A   & P & R & F1 & A & P & R & F1 \\
 & 87.88\% & 1 & 70\% & 82.35\% & / & / & 73.81\% & /  \\ \bottomrule
\end{tabular}%
\vspace{-2ex}
\end{table}

\noindent\textbf{Results.}
Table~\ref{tab:metrics} presents the performance of ATPChecker in identifying two aspects: i) whether host apps accurately declare the usage of TPLs in their privacy policies, referred to as "TPL list", and ii) whether host apps accurately declare the data usage of TPLs in their privacy policies, denoted as "TPL data". The metrics used for evaluation include Accuracy (A), Precision (P), Recall (R), and F1-score.
The evaluation assesses ATPChecker's performance from two perspectives: Trace Level, which measures the number of metrics based on the behaviors identified by ATPChecker in each app, and App Level, which counts the metrics based on the number of apps.
For instance, suppose ATPChecker identifies that an app conceals the usage of two TPLs in its privacy policies and the app indeed conceals the usage accurately. In this case, we record TP (True Positive) as 2 for Trace Level in TPL List and count TP as 1 for App Level in TPL list.

Table~\ref{tab:metrics} reveals that ATPChecker achieves an accuracy of over 93.29\% at the trace level and 87.88\% at the app level for identifying the declaration of TPL usage in host apps' privacy policies.
As for the false negatives related to ATPChecker's ability to identify the declaration of TPL usage in host apps' privacy policies, ATPChecker utilizes keyword matching in the compliance identification modules. However, mismatches may occur if the TPL's name coincides exactly with certain words within the sentence.
Although ATPChecker cannot identify the false positives for declaration of host apps' data sharing with TPLs in their privacy policies, ATPChecker still achieves 78\% recall at trace level and 73\% recall at app level.
The occurrence of false negatives primarily arises in cases where host apps make data-sharing claims within lengthy sentences, exceeding 70 words. In such instances, ATPChecker may incorrectly break down the subject, predicate, and object, leading to inaccurate identification of the subject involved in the data-sharing behaviors.

Table~\ref{tab:metrics} reveals that ATPChecker identifies 56 out of 254 (22\%) host apps, i.e., True Positives for identifying declaration of TPL usage in host apps' privacy policies at app level, do not clearly claim their TPL usage in privacy policies and 62 out of 254 (24\%) host apps, i.e., True Positives for identifying declaration of data sharing with TPLs in host apps' privacy policies, do not clearly claim the data sharing with TPLs in host apps' privacy policies.
This situation may arise due to the lack of awareness among the developers of these host apps regarding the usage of TPLs, such as certain development tools or advertising libraries, integrated within their apps. It is also possible that these developers may not have specifically recognized the extent of data sharing that occurs with TPLs.

\answer{4}{
ATPChecker identifies that more than 22\% of apps fail to comply with regulatory requirements by not disclosing their TPL usage in their privacy policies. Additionally, it identifies that over 20\% of apps violate regulatory requirements by not adequately disclosing their data interactions with TPLs.
}

\section{Discussion}

%

This section discusses the limitation of {\oursystem}, the discussion of assumption, threats to validity, and future work.

\subsection{Limitations}
{\oursystem} was designed to identify whether TPLs' personal information usage complies with regulations and whether host apps' TPL usage complies with regulations.
%
{\oursystem} can only assess the binary files of TPLs and their corresponding privacy policies if they were collected simultaneously. It cannot guarantee that the binary files and privacy policies are of the same version.
Note that TPLs' functions may be changed with the updates of the TPL versions.
It is possible to trace TPL versions from {\maven}, but it is not easy to trace the privacy policy of corresponding versions.

{\oursystem} is based on static analysis tools like soot and flowdroid.
{\oursystem} cannot handle some dynamic behaviors of TPLs (e.g., reflection, dynamic class loading), and it also cannot process large apk~\cite{arzt2014flowdroid} files due to the limited memory space. 
Furthermore, {\oursystem} can neither handle host apps in the format of \textit{xapk} which is created by Apkpure~\cite{apkpure} nor process native libraries.

Moreover, {\oursystem} is limited to the pre-defined patterns and string matching for identifying the statements and TPL usage in collected privacy policies. 
%
The string matching method may result in some incorrect matches, for example, some matches contain the keyword but trace does not have the function of data collection.
%
{\oursystem} will fail to detect those issues if adversaries use novel patterns to hide their data usage statements~\cite{boucher2022bad, zhao2021structural} .
%

\subsection{Discussion of assumption}

{\oursystem} was designed to identify the compliance issues between TPLs' behavior and privacy policies, and host apps' TPL usage and privacy policies.
TPLs are mainly used as an additional part of apps to enhance app functionalities and may not be responsible for data access behaviors or work in the role of data controllers as defined in GDPR. 
However, we cannot assume all TPLs only conduct user data access behavior for assisting apps. 
Existing work~\cite{tangtse:21a, meng2016price, liu2016alde} has demonstrated that development tool TPLs, such as firebase~\cite{tangtse:21a}, may leak users’ privacy without developers' consciousness.
Besides, regulations, such as GDPR Article 35 and SGSDK Article 5.3, request clearly giving the purpose of data processing. 
Even if we could assume TPLs only access user data for specific functionalities and would never share or collect the data, we still recommend TPLs provide privacy policies to clearly state their user data access behavior and related purpose.
It can help not only app developers better understand TPL's functionalities but also TPL developers avoid legal disputes.
Thus, we assume all TPLs should provide privacy policies when evaluating {\oursystem}.

\subsection{Threats to validity}
The first threat comes from the language used in privacy policies, it is not trivial to identify and switch the language of privacy policy during the collecting phase, which may lead to a missing collection of some privacy policies.
%
As TPLs or apps are published in different countries, such as China or Korea, the default language used in the privacy policy website is provided using their native language. 

Another threat comes from inconsistent versions of TPLs and their privacy policies.
Notice that all our data was collected from Feb-2022 to Apr-2022.
There may be cases where the TPLs' or apps' functions have been updated, but the privacy policies have not been updated in time. 
This can lead our system to misidentify that the behavior of the software is inconsistent with the privacy policy and violates regulation requirements.
We will mine software and privacy policy version issues in future work.

Moreover, the third threat is due to the lack of a large-scale labeled data set.
We only crawl the TPL list and the top 10 host apps from {\appbrain}.
It is very laborious to label privacy policies and collect the ground truth of data usage in software and privacy policies.

\subsection{Future work}
We will equip {\oursystem} with the capability of analyzing the data collection purpose, because such information can help researchers better understand the code and detect violations. 
%
%
%
%
Furthermore, writing legal privacy policies remains challenging and time-consuming work for TPL developers.
Automatic privacy policy generation methods are in urgent need.
%
Although there are privacy policy generation methods~\cite{le2015ppg, zimmeck2021privacyflash} for apps, those methods are not suitable for generating privacy policies for TPLs.
In future work, we will develop automatic TPL privacy policy generation methods by combing regulations requirements analysis~\cite{zhSDKLaw, gdpr, ccpa}, natural language processing methods~\cite{zhao22ca4p} and TPL analysis techniques~\cite{atvhunter2021ICSE}.

\section{Related Work}




\noindent\textbf{Privacy policy conflict identification.} 
XFinder~\cite{wang2021understanding} identifies the cross-library data harvesting in Android apps with dynamic analysis.
XFinder identifies third-party libraries' usage by comparing the caller's and callee's package names.
XFinder also restores reflection invocations using two predefined patterns.
For conflict identification, XFinder manually parses the term-of-service of 40 TPLs and then uses NLP techniques to extract data sharing policies.
Nguyen et al.~\cite{nguyen2021share} investigate whether apps achieve users' consent before sharing personal information.
The authors use dynamic analysis to identify the network traffic and data sharing behaviors.
They determine whether the shared data are identifiable personal data by comparing the same traffic collected from different times or the same traffic from different devices.
The ablation experiments are designed to determine whether the data-sharing action achieves users' explicit consent.


PAMDroid~\cite{zhang2020does} analyzes the impact of misconfigurations of analytic services in Android.
After analyzing 1000 popular apps, PAMDroid finds 52 of 120 apps misconfigure the services and lead to a violation of either the service providers' term-of-service or the app's privacy policy.
%
%
PPChecker~\cite{ppchecker} detects the conflicts in apps' privacy policies, but only determines whether apps' privacy policies provide TPLs' privacy policy links and interactions of five permission related personal information  with 81 TPLs. 
POLICHECK~\cite{ppcheck} identifies the app's data sharing with third parties using dynamic analysis. 
%
POLICHECK finds that 49.5\% of apps disclose their third-party sharing practices using vague terms and 31.1\% of data flows as omitted disclosures.
Existing works ignore analyzing whether TPLs satisfy the regulation of requirements.
%




\noindent\textbf{TPL data leakage identification.} Razaghpanah et al.~\cite{razaghpanah2018apps}  detect third-party advertising and tracking services using dynamic analysis.
They use dynamic analysis to identify the advertising and tracking services.
Specifically, they use a free  {\app}, namely Lumen Privacy Monitor, to collect all network traffic generated by all {\app}s installed on the device.
With limitations of Lumen, the proposed system can only identify limited personally identifiable information and unique identifier, e.g., IMEI.
%
He et al.~\cite{he2018dynamic} use dynamic analysis to analyze the invocation path between predefined source and sink to identify the privacy leakage of third-party libraries.
Their system only concentrates on Android permission-related personal information.
Their experiments on 150 popular {\app}s demonstrate that their proposed dynamic methods achieve real-time detection and 97.4\% accuracy.
Ekambaranathan et al.~\cite{ekambaranathan2021money} concentrate on children's {\app}s data usage and disclosure.
The researchers conduct surveys and interview with {\app} developers to understand why {\app}s disclose children's personal data.
%
Liu et.al.~\cite{liu2019privacy} analyze the data leaking of nine analytics libraries in 300 apps.
They conduct static and dynamic analyses to mitigate the privacy risk caused by analytics libraries.%

\section{Conclusion}

We propose an automatic third-party library regulation compliance checker, namely {\oursystem}.
%
{\oursystem} was designed to identify whether TPLs satisfy regulation requirements, i.e., whether TPLs provide privacy policies and correctly claim their data usage, and whether host apps correctly disclose their usage and data interaction with TPLs.
{\oursystem} discovered that over 23.4\% TPLs incorrectly provide the privacy policies, 37\% TPLs do not disclose all of their data usages, and over 65.64\% apps miss disclosing their personal information interaction with TPLs.
%
%
%

\section{Data Availability}

We make our dataset and tool publicly available to facilitate research in this area.
We release the code and data to other researchers by responsibly sharing a private repository. 
The project website with instructions to request access is at:
\url{https://doi.org/10.5281/zenodo.7932665}.
%
Besides,  our data set is constructed by gathering publicly available privacy policy websites and apps without posing any ethical problems.

\section{Acknowledgment}
We thank the anonymous reviewers for their helpful comments. 
This work was partially supported by Hong Kong RGC Projects (No. PolyU15219319, and No. PolyU15224121), HKPolyU Start-up Fund (BD7H), and National Natural Science Foundation of China (No. 62202406). Research Grant from Huawei Technologies Co., Ltd.

%

\bibliographystyle{IEEEtran}
\bibliography{bib}


\end{document}